# Scale free networks of superconducting striped grains tuned at Fano resonances for high $T_c$

**Antonio Bianconi**

[1]*Rome International Center for Materials Science Superstripes RICMASS Via dei Sabelli 119A, 00185 Roma, Italy*

[2]*Mediterranean Institute of Fundamental Physics, Via Appia Nuova 31, 00040 Marino Italy*

[3]*Department of Physics, Sapienza University of Rome, P.le Aldo Moro 2, 00185 Roma*

E-mail: antonio.bianconi@ricmass.eu

**Abstract.** Controlling lattice complexity or inhomogeneity of the doped oxide superconductors is shown to be a key term for control of the superconductivity critical temperature. All high temperature superconductors show anisotropic gaps and/or multi-gap superconductivity in the "clean limit". They show the shape resonance in superconducting gaps that is a type of Fano resonance driven by the Josephson-like term for pairs transfer. These resonances occur where one of the multiple Fermi surface spots is near a 2.5 Lifshitz transition. The Fano antiresonance occurs between a BCS condensate and a BEC-like condensate. In a system near a 2.5 Lifshitz transition the lattice is close to an instability. This drives to the formation of a granular superconducting phase made of a charge density metal formed by puddles of ordered oxygen interstitials or ordered local lattice distortions (static short range charge density waves) detected by scanning nano-x-ray diffraction imaging. We show that the scale free networks of superconducting grains tuned to a Fano resonance is an essential feature of high temperature superconductors that favours high $T_c$.

## 1. Introduction

Understanding the mechanism that allows a quantum condensate to resist to the decoherence attacks of temperature is a major fundamental problem of condensed matter. The Bardeen Cooper Schrieffer (BCS) wave-function [1-6] of the superconducting ground state has been constructed based on the theory of configuration interaction of all electron pairs (+k with spin up, and -k with spin

.

down) on the Fermi surface in an energy window called the energy cut off of the interaction,

$$|\Psi_{BCS}\rangle = \prod_k (u_k + v_k c^+_{k\uparrow} c^+_{-k\downarrow})|0\rangle \qquad (1)$$

where $|0\rangle$ is the vacuum state, and $c^+_{k\uparrow}$ is the creation operator for an electron with momentum k and spin up. The many body BCS condensate with off-diagonal long range appears at a gap energy $\Delta(\kappa)$ below the Fermi level. The Schrieffer idea [7] came from the configuration interaction theory by Tomonaga involving a pion condensate around the nucleus [8]. The k-dependent structure of the interaction gives different values for the gap $\Delta(k)$ in different segments of the Fermi surface. The superfluid order parameter i.e., the superconducting gap superconductivity is anisotropic in the k-space i.e., different in different locations of the k-space due different pairing strength. The k-dependent gap equation is given by

$$\Delta(\mu,k) = -\frac{1}{2N}\sum_{k'} \frac{V(k,k')\Delta(k'_y)}{\sqrt{(E(k')+\varepsilon_{k'}-\mu)^2 + \Delta^2(k'_y)}} \qquad (2)$$

and the critical temperature is given by

$$\Delta(k) = -\frac{1}{N}\sum_{k'} V(k,k') \frac{tgh(\frac{\xi(k')}{2T_c})}{2\xi(k')}\Delta(k') \qquad (3)$$

where the $\xi_n(k)=\varepsilon_n(k)-\mu$.

These original BCS formula describe the anisotropic superconductivity in the "clean limit", where the single electron mean free path is larger than the superconducting coherence length. Real superconducting materials [9,10] have impurities and lattice disorder therefore the condition for the mean-free path $l > hv_F/\Delta_{av}$ where $v_F$ is the Fermi velocity and $\Delta_{av}$ is the average superconducting gap was considered to be impossible to be satisfied [9,10]. In fact it is a very strict condition that implies that the impurity scattering rate $\gamma_{ab} << (1/2)(K_B/\hbar)T_c$ should be smaller than few meV. The 40 years period from 1960 to 2000 was dominated by the "dirty limit dogma" [11,12] assuming an effective single Fermi surface for each superconductor. This dogma stated that all metals are in the "dirty limit" since impurity scattering and hybridization always mixes the wave functions of electrons on different spots in the same Fermi surface and on different Fermi surfaces due to multiple bands crossing the Fermi level. The "dirty limit" dogma justifies the approximation in the theory of a k-independent V(k,k') interaction i.e., an isotropic pairing interaction that is assumed to be a constant, $V_0$ [11,12]. This approximation allows to derive the approximated simple universal BCS formula for the critical temperature $T_C$ related to a single superconducting energy gap $\Delta_0$ the energy needed to break the cooper pairs, averaged over many k-points in different bands:

$$\frac{2\Delta_0}{K_B T_C} = 3.52 \qquad (4)$$

$$\frac{T_c}{T_F} = \frac{0.36}{k_F \xi_0} \quad (5)$$

$$K_B T_c \propto \hbar \omega_0 e^{-1/\lambda_{eff}} \quad (6)$$

where $T_F$ is Fermi temperature, $k_F = 2\pi/\lambda_F$ is the wave-vector of electrons at the Fermi level, $\xi_0$ is the coherence length of the condensate, related with the size of the pair, $\hbar\omega_0$ is the energy if the boson meditating the pairing interaction, the effective coupling term $\lambda_{eff} = V_0 N_{tot}$ is the product of the pairing strength $V_0$ and the total density of states (DOS) at the Fermi level $N_{tot}$. Moreover using this approximation the BCS theory predicts that the critical temperature depends on isotope substitution with a power law dependence $T_c \propto M^{-\alpha}$, where M is the atomic mass with the isotope coefficient α=0.5 [9,10]

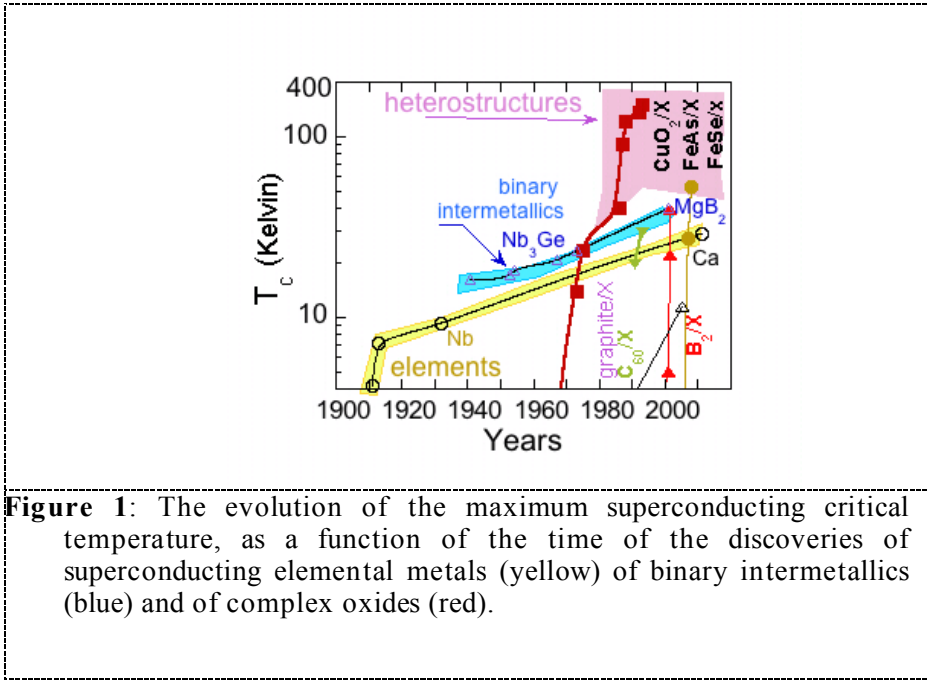

**Figure 1**: The evolution of the maximum superconducting critical temperature, as a function of the time of the discoveries of superconducting elemental metals (yellow) of binary intermetallics (blue) and of complex oxides (red).

These popular formulas are simple but they have completely washed out all interesting possible effects on the macroscopic superconducting parameters due to quantum interference effects in the configuration interaction between pairing channels in different points of the k-space the "clean limit". According with De Gennes this is not a problem since no spectacular effects have to be expected from the k-dependent V(k,k') interaction [12]. The last 40 years of the XX century can be called the period of the "dirty limit" dogma. The road map for the research of new high $T_c$ materials based on the BCS approximated formulas was focused on metals showing peaks of total DOS, strong electron phonon coupling, and high energy phonon modes. The details of superconductivity in different materials have been described by the Migdal-Eliashberg approximation including the details if the attractive electron-phonon interaction and the Coulomb repulsion, keeping the assumption that the energy scale of the pairing interaction is much smaller than the Fermi energy $\hbar\omega_0 \ll E_F$ called the so called "adiabatic limit" i.e. the chemical potential in the metallic material is assumed to be far away from band-edge. The materials enter in the antiadiabatic regime $E_F \approx \hbar\omega_0$ and the charge carriers become polarons. Moving the chemical potential toward a band edge where the system approach a metal-to-insulator transition. In the extreme antiadiabatic regime $E_F \ll \hbar\omega_0$ at the metal-to-insulator transition the BCS condensate become a BEC (Bose Einstein Condensate) that below $T_c$ is formed below the band edge. Therefore near a band edge the system is in the so called BEC-BCS crossover [4-6].

The materials science research has made a slow progress in the search for new high temperature superconductors in these last 100 years as shown in Fig. 1. The main discontinuities occurred by shifting from simple elemental metals to binary intermetallics in the 30's and the record was reached in binary alloys in metastable phases near a lattice instability [10]. Finally in 2001 the highest $T_c$ in a binary intermetallics has been reached in $MgB_2$ that is a multi-gap superconductor in the "clean limit" and the chemical potential is driven near a band edge i.e. , below the top of the boron $2p_{x,y}$ sigma band [11]. This key discovery of material science has shown that both the "dirty limit dogma" and the adiabatic approximation in BCS theory need to be abandoned in the regime of high temperature superconductivity.

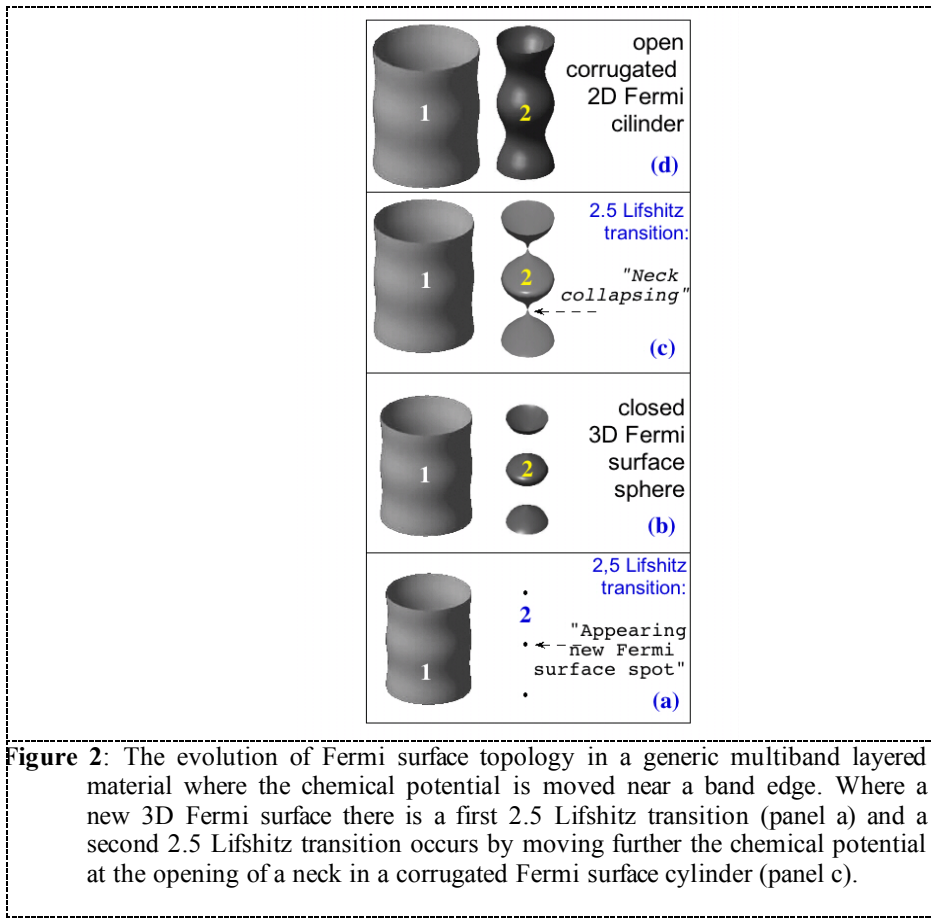

**Figure 2**: The evolution of Fermi surface topology in a generic multiband layered material where the chemical potential is moved near a band edge. Where a new 3D Fermi surface there is a first 2.5 Lifshitz transition (panel a) and a second 2.5 Lifshitz transition occurs by moving further the chemical potential at the opening of a neck in a corrugated Fermi surface cylinder (panel c).

The same scenario emerges for the superconductivity in cuprates where it was quickly clear that they show a polaronic liquid [14] in the antiadiabatic limit by using local and fast structural probes: extended x-ray absorption fine structure (EXAFS) [15] and x-ray absorption near edge structure [16]. It was shown that local lattice distortions in cuprates [17] form short range puddles of charge and orbital density waves. In each puddle the lattice modulation gives multiple subbands crossing the Fermi level resulting in multiple Fermi surface arcs with heterogeneous spots in the k-space [18,19]. Recently the presence of multiband superconductivity in proximity of band edges clearly appear in electron doped iron based superconductors in Ba(Fe$_{1-x}$Co$_x$)$_2$As$_2$ [20] and a(Fe$_{1-x}$Ni$_x$)$_2$As$_2$ [21] where the Tc dome occur in the proximity of 2.5 Lifshitz transitions [22] a predicted by Innocenti et al [23]

## 2. The standard multiband BCS theory

Multiband superconductivity [24-25] appears in anisotropic superconductivity, where the gaps are different in different Fermi surfaces. The multiple Fermi surfaces generated by multiband cannot be reduced to a single effective band since the particles remain distinguishable because they have different symmetry and/or are located in different spatial portions of the material. In this scenario the single particle hopping in presence of impurities and hybridization are forbidden. For example the condensate many body BCS wave-function for a two band superconductor made of a first *a*-band and a second *b*-band is given by:

$$\left|\Psi_{Kondo}\right\rangle = \prod_k (u_k + v_k a^+_{k\uparrow} a^+_{-k\downarrow}) \prod_{k'} (x_{k'} + y_{k'} b^+_{k'\uparrow} b^+_{-k'\downarrow}) \left|0\right\rangle \quad (7)$$

The term corresponding to the transfer of a pair from the "a"-band to the "b"-band and viceversa appears with the negative sign [26] in the expression of the energy.

$$\sum_{k,k'} J(k,k')(a^+_{k\uparrow} a^+_{-k\downarrow} b_{-k\downarrow} b_{k\uparrow}) \quad (8)$$

where $a^+$ and $b^+$ are creation operators of electrons in the "a" and "b" band respectively and $J(k,k')$ is an exchange-like integral. This gain of energy is the origin of the increase of the transition temperature driven by this exchange-like [26] interaction between pairs in different points in the k-space. This is called also a Josephson-like pairing term, in fact the Josephson effect [27] describes the transfer of pairs between two different condensates in different spatial positions. This Josephson-like pairing interaction appears in the standard BCS multiband theory with a square exponent therefore it may be repulsive as it was first noticed by Kondo [26]. It is therefore different from the Cooper pairing process that is the conventional attractive intraband attraction for electron at the Fermi level in the single band BCS theory. In the case of repulsive Josephson-like interaction in multiband superconductivity the order parameter shows the sign reversal between the different bands or different points in the k-space. This became popular in the case of iron based superconductors with the name of $s\pm$ mechanism.

In anisotropic superconductivity of a multiband superconductor the k-dependent gaps in each band n depend on the gaps in other bands

$$\Delta_n(\mu, k_y) = -\frac{1}{2N} \sum_{n', k_y', k_x'} \frac{V_{n,n'}(k, k') \Delta_{n'}(k_y')}{\sqrt{(E_{n'}(k_y') + \varepsilon_{k_x'} - \mu)^2 + \Delta^2_{n'}(k_y')}} \quad (9)$$

where the k-dependent coupling is given by

$$V^o_{n,k_y;n',k_y'} = -V_o \int_S dx\, dy\, \psi_{n,-k}(x,y) \psi_{n',-k'}(x,y) \psi_{n,k}(x,y) \psi_{n',k'}(x,y) \quad (10)$$

### 3. A type of Fano resonance: Shape resonances in superconducting gaps

In the standard theory of multiband superconductivity the Fermi level is assumed to far away from each band edge: i.e., the intraband pairing in each band is in the adiabatic limit. However some authors [23,28-31] have proposed that spectacular effects similar to a Fano resonance [32] occur in the superconductivity mechanism in a multiband superconductor when the chemical potential is tuned near a band edge at a 2.5 Lifshitz transition. In this case the intraband pairing mechanism enters in the antiadiabatic limit in the particular band where the Fermi level is near the band edge.

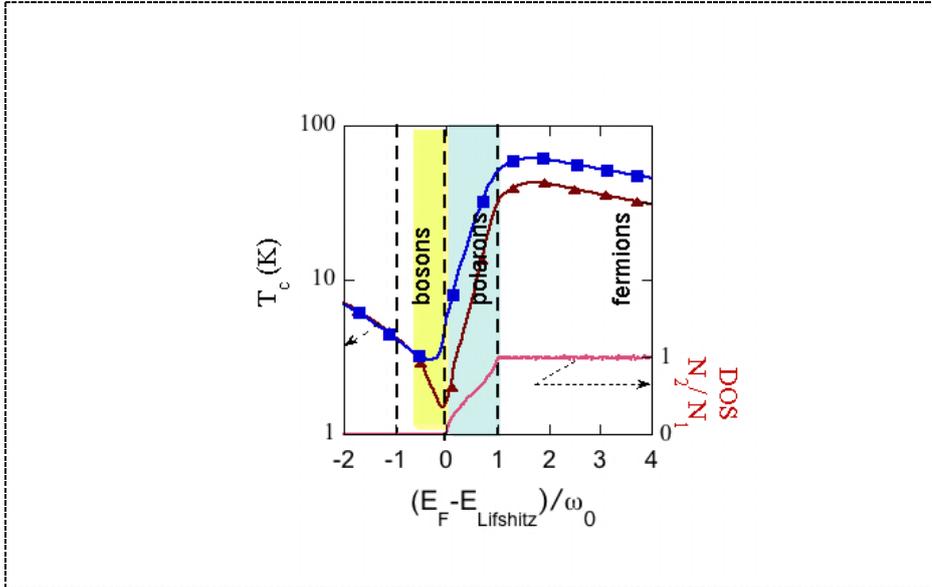

**Figure 3**: The critical temperature of a layered multi-band superconductor tuning the chemical potential near a 2.5 Lifshitz transition where the Fermi surface of a new appearing band evolves as in Fig. 2. The critical temperature in log-scale is plotted as a function of the Lifshitz energy parameter $z = (E_F - E_{Lifshitz})/\omega_0$ where $\omega_0$ is the energy of the attractive intraband pairing interaction. The plot show two cases of Josephson-like pair exchange coupling ratio: a) $c12/c11=-1.08$ (*blue filled squares*) and $c12/c11=-0.43$ (*black filled triangles*). The colored regions indicate the BEC regime of the condensate (yellow) $-1<z<0$ and bipolaronic condensate (blue) $0<z<1$ in the new appearing Fermi surface.

The system can be treated in the BCS scheme [27,31] if the equation for the critical temperature near a 2.5 Lifshitz transition is solved together with the density equation as suggested by Leggett [5] taking into account that in this regime there is a large variation between the chemical potential in the superconducting phase and the normal phase

$$\Delta_n(k) = -\frac{1}{N}\sum_{n'k'} V_{n,n'}(k,k') \frac{tgh(\frac{\xi_{n'}(k')}{2T_c})}{2\xi_{n'}(k')} \Delta_{n'}(k') \quad (11)$$

$$\rho = \frac{1}{S}\sum_{n}^{N_b}\sum_{k_x,k_y}(1 - \frac{\varepsilon_n(k_x,k_y) - \mu}{\sqrt{(\varepsilon_n(k_x,k_y) - \mu)^2 + \Delta_{n,k_y}^2}}) \quad (12)$$

We consider a two band superconductor where there is first large 2D cylindrical Fermi surface and the chemical potential is tuned across a 2.5 Lifshitz transition so that the Fermi surface topology of the new appearing Fermi surface evolves as shown in Fig. 2. The superconducting critical temperature as a function of the Lifshitz energy parameter measuring the energy distance form the 2.5 Lifshitz transition for the appearing of the a new Fermi surface is plotted in Fig. 3. This figure provides a typical example of shape resonance in the superconducting gaps. The chemical potential crosses the first 2.5 Lifshitz transition, for the appearing of the new Fermi surface spot, at the value of the Lifshitz energy parameter $z = 0$, and the second 2.5 Lifshitz transition "opening a neck" at $z = 1$. The coupling in the first Fermi band is assumed to be in the standard BCS weak coupling regime $c11 = 0.22$. We consider here the case of strong coupling in the second band where the attractive coupling term $c22$ is about two times larger than the Cooper pairing coupling parameter in the first band $c22/c11 = 2.17$. When the Lifshitz energy parameter is in the range $-1<z<0$ a *BEC condensate* is formed in the second band since all charges in the second band condense. At $z = 0$, we show evidence for the antiresonance due to negative interference effects between different pairing channels in the BCS condensate in the first band and the BEC in the new appearing band. This clearly shows is the quantum interference nature of shape resonances. The second curve shows the higher critical temperature for a larger $c12$ Josephson-like pair transfer value. *T*c reaches the minimum value at the antiresonance for $z = 0$ in the limit of weak Josephson-like coupling, while for a strong Josephson-like coupling $c_{12}$ $c_{21}$ the minimum due to the Fano-like antiresonance approach $z=-1$. Clearly we show in Fig. 3 that the shape resonance [23,28-31] as a similar line-shape as a Fano resonance [32].

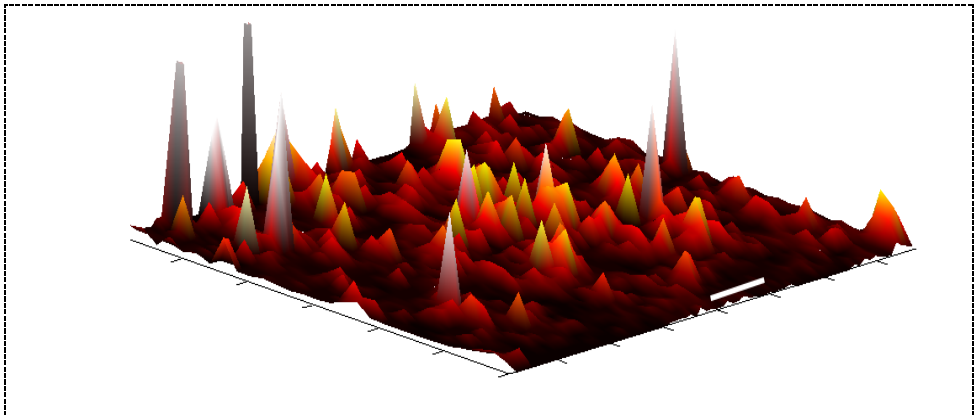

**Figure 4**: The landscape of nanoscale striped puddles of local lattice distortions in a cuprate superconductor $La_2CuO_{4+y}$ measured by scanning nano x-ray diffraction. The position dependence of the Q3-LLD superstructure intensity in the $La_2CuO_{4+y}$ crystals with critical temperature, $T_c$, 37 K is plotted. The white bar indicates a length of 100 microns. The scanning XRD images show the better self organization of LLD droplets favoring higher $T_c$.

These results provide a roadmap for material design of new room temperature superconductors with a lattice geometry that allow
a) a multi-band superconductor in the clean limit with several bands crossing the Fermi level;
b) the different bands crossing the Fermi energy should have *different parity* and *different spatial locations* to avoid hybridization

c) the chemical potential is tuned near a 2.5 Lifshitz transition near a band edge.
d) the exchange-like or Josephson-like pair transfer integral should be as large as possible
e) a first condensate is in the adiabatic regime and a second condensate in the anti-adiabatic regime
f) The chemical potential can be tuned by pressure or gate voltage to get the Fano shape resonance in the superconducting gaps between pairing scattering channels in a BCS-BEC crossover
g) Tune the chemical potential by gate voltage tuning of using illumination or pressure so that the bipolaron condensate is converted into a BCS condensate in fact the maximum $T_c$ occurs at a Lifshitz energy parameter z near 1.5 -2.

**4. Optimum lattice inhomogeneity favoring high $T_c$.**

A characteristic feature of cuprate superconductors is the lattice complexity. In the oxides of copper the defect order is correlated with their high superconducting transition temperatures.

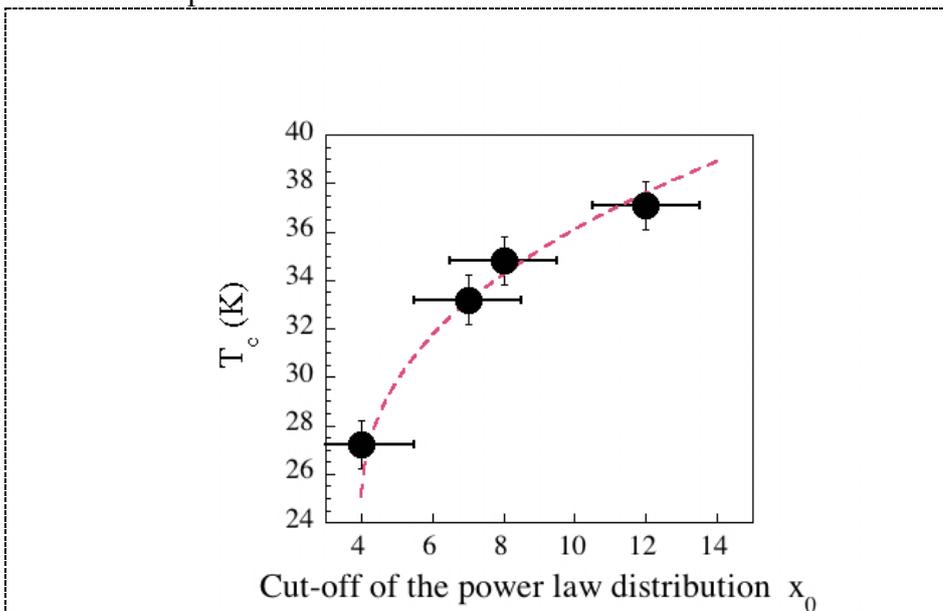

**Figure 5**: The critical temperature $T_c$ in the range 25 K < $T_c$ < 37 K for several super oxygenated La214 samples is plotted as a function of the cut-off parameter of the distribution of the LLD droplets density probed by the intensity distribution of the Q3-LLD superstructure satellites. Error bars in the critical temperature are of ± 1 K. The dashed line is the fit with a power law curve with exponent 0.4 ± 0.05, in agreement with theoretical predictions in references [38-39] for granular superconductivity on a scale invariant network

The oxygen interstitials can be very inhomogeneous, even in "optimal" superconducting samples with a scale free distribution favouring high $T_c$ as recently shown by scanning nano X-ray diffraction [33]. The oxygen interstitials self organization can be controlled by x-ray illumination [34].

Indeed, in a sample where a the oxygen interstitials are frozen in a random distribution at a temperature lower than 200K, the local lattice distortions detected by EXAFS and XANES [17-19] are shom to get self organized formimg short range striped puddles [35,36]. Moreover using a x-ray nanobeam of 300 nm in diameter, we have obtained the imaging of the regions in $La_2CuO_{4+y}$, that contain incommensurate modulated local lattice distortions (LLD). The puddles are determined by self organization of local lattice distortions with superlattice wave-vector **q3** = 0.21 **b***+0.29 **c** therefore are called with the acronym Q3-LLD [37]. Fig. 4 shows the spatial distribution of Q3-LLD puddles for a superconducting sample. This result shows that heterogeneous phases of nanoscale phase separation is common to cuprates and iron based superconductors [38].

The Q3-LLD droplets form networks whose nature varies with superconducting critical temperature. We have used X-ray micro-diffraction apparatus at the ESRF to map the evolution of the Q3-LLD satellites for four single crystals of electrochemically doped $La_2CuO_{4+y}$, from the underdoped state to the optimum doping range, $0.06<y<0.12$.

The probability distributions, of the Q3-LLD XRD intensity for single crystals of electrochemically doped $La_2CuO_{4+y}$ from the underdoped state to the optimum doping range, $0.06<y< 0.12$ follow a power law distribution $P(x) \propto x^{-\alpha} \exp(-x/x_0)$ with a variable exponential cut-off $x_0$ with a constant power-law exponent $\alpha = 2.6 \pm 0.1$.

In Fig. 5 we have plotted the critical temperature of samples in the range 27-38 K associated with the droplet network of Josephson coupled nano-grains, as a function of the cut-off of the probability distribution $x_0$ of the intensity of x-ray intensities due to the variable density of Q3-LLD puddles. The critical temperature scales with the cut-off according to a power law with an exponent 0.4 ± 0.05. This result points again toward the importance of connectivity and an optimum inhomogeneity for high critical temperature. They are in qualitative agreement with the theoretical prediction of the increase of $T_c$ in a granular superconductor on an annealed complex network made of Josephson coupled grain following a power law distribution with a finite cut-off [38.39]. In fact, for a power law distribution of links in a granular superconductor with an exponent α=2.6, the critical temperature is predicted to increase as a function of the cut-off with an exponent 3–α, as observed experimentally supporting the theory of quantum phase transitions in network of superconducting grains [39,40] and for bosonic scale free networks. [41]

5. Conclusions

We have shown that all high temperature superconductors show k-dependent pairing anisotropic and multigap superconductivity. The high $T_c$ dome appears in the proximity of 2.5 Lifshitz transition near a band edge showing that the electrons in one of the multiple Fermi surfaces are in the crossover BCE-BCS regime. At list one of the limit of the high $T_c$ dome is determined by the antiresonance regime of the shape resonance in the superconducting gaps where $T_c$ drops toward zero at the 2.5 Lifshitz transition for the appearing of a new Fermi surface. At this particular regime there is a BEC-BCS crossover with coexisting BEC and BCS condensates that gives an anti-resonance in the configuration interaction controlled by the Josephson-like pair transfer term. This term gives an increasing high Tc amplification when the BEC is converted to a polaronic pair condensate and finally the maximum Tc is reached where the polaronic condensate moves toward a BCS condensate.

The materials near a 2.5 Lifshitz transition are in a metastable phase due to lattice instability and show networks of striped superconducting grains. It seem to us that the maximum $T_c$ is reached at the particular condition where in each grain with a striped structure the critical temperature is amplified by the shape resonance in the superconducting gaps and the grains form a scale free network where $T_c$ is pushed to the highest possible temperature.